\documentclass{mem}
\usepackage{natbib}\usepackage{txfonts}\usepackage{balance}
\usepackage{graphicx}
\usepackage[bookmarks,colorlinks,citecolor=cyan,breaklinks]{hyperref}
\idline{75}{282}
\newcommand\actaa{{Acta Astronomica}}%

\begin{document}
\def\teff{$T\rm_{eff }$}
\def\hst{{\it HST\/}}
\def\kms{$\mathrm {km s}^{-1}$}
\def\ngc#1{\hbox{NGC\,#1}}
\def\feh{\hbox{\rm [Fe/H]}}
\def\afe{\hbox{\rm [$\alpha$/Fe]}}

\title{ The globular cluster \ngc6528 the ferrous side of the Galactic Bulge} 
   \subtitle{}

\author{
 E.~P. Lagioia\inst{1},
 A.~P. Milone\inst{2},
 G. Bono\inst{1,3}, 
 P.~B.~Stetson\inst{4},
 A.~Aparicio\inst{5,6},
 R.~Buonanno\inst{1,7},
 A.~Calamida\inst{8},
 M.~Dall'Ora\inst{9},
 I.~Ferraro\inst{3},
 R.~Gilmozzi\inst{10},
 G.~Iannicola\inst{3},
 N.~Matsunaga\inst{11},
 M.~Monelli\inst{4,5},
 P.~G.~Prada Moroni\inst{12}
 \and
 A.~Walker\inst{13}
 }

  \offprints{E.~P. Lagioia}

\institute{ 
Dipartimento di fisica, Universit\`a degli studi di Roma -- Tor Vergata, Via
della Ricerca Scientifica 1, 00133, Roma, Italy \email{eplagioia@roma2.infn.it}
\and 
Research School of Astronomy and Astrophysics, The Australian National
University, Cotter Road, Weston, ACT, 2611, Australia;
\email{milone@mso.anu.edu.au}
\and 
INAF -- OAR, Via Frascati 33, 00040 Monte Porzio Catone, Italy
\and 
Dominion Astrophysical Observatory, Herzberg Institute of Astrophysics, National
Research Council, 5071 West Saanich Road, Victoria, BC V9E 2E7, Canada
\and 
Instituto de Astrof\`isica de Canarias, E-38200 La Laguna, Tenerife, Canary
Islands, Spain
\and 
Dept. Astroph., Univ. of La Laguna, 38200 La Laguna, Tenerife, Canary Islands, Spain 
\and 
INAF -- OACTe, via M. Maggini, 64100 Teramo, Italy
\and 
Space Telescope Science Institute, 3700 San Martin Dr., Baltimore, MD 21218, USA
\and 
INAF -- OAC, Salita Moiariello 16, 80131 Napoli, Italy
\and 
ESO, Karl-Schwarzschild-Stra{\ss}e 2, 85748, Garching, Germany
\and 
Kiso Observatory, Institute of Astronomy, School of Science, The University of
Tokyo, 10762-30, Mitake, Kiso-machi, Kiso-gun,3 Nagano 97-0101, Japan
\and 
Dipartimento di Fisica, Universit\`a di Pisa, Largo E. Fermi, 56127 Pisa, Italy
\and 
CTIO, National Optical Astronomy Observatory, Casilla 603, La Serena, Chile
}

\authorrunning{E.~P. Lagioia}

\titlerunning{The globular \ngc6528 the ferrous side of the Galactic Bulge}

\abstract{
We present new and accurate optical photometry of the Bulge globular cluster 
\ngc6528. The images were collected with ACS at HST and together with WFC3 
(UVIS, IR) allowed us to measure the proper motion to separate cluster 
and field stars. 
We adopted two empirical calibrators and we found that \ngc6528 is coeval 
with and more metal-rich than 47\,Tuc. Moreover, it appears older and more 
metal-poor than the super-metal-rich old open cluster \ngc6791. 
We also performed a preliminary analysis of field stellar populations located 
around \ngc6528 and \ngc6522 by using ground-based near-infrared photometry 
collected with SOFI at NTT. The comparison of evolved stellar components 
(red giant branch, red horizontal branch, red clump stars) indicates that 
they share similar properties in this region of the Baade's Window.   

}

\maketitle{}

\section{Introduction}
The Bulge is a complex component of the Galactic spheroid whose stellar content
shows a broad range of kinematic and chemical components. Different
observational campaigns connected either with microlensing events in the
direction of the Galactic center (OGLE, EROS, MACHO,
\citep{Uda94,Zha95,Alc95,Aub93}, or with near-infrared (NIR) integrated surface
photometry from COBE/DIRBE observations \citep{Dwe95}, or with star counts from
the 2MASS survey \citep{Lop05}, have found evidence that the Bulge has a
boxy/peanut morphology, with the long axis lying onto the Galactic plane at an
angle of $\phi_0\simeq 20^\circ$ from the Sun-Galactic center direction
\citep{Rat07,Ger02}. 
More recent investigations based on resolved stellar populations have also found
evidence of a bar by analyzing the (bimodal) distribution of the Red Clump (RC)
stars in the NIR color-magnitude diagram (CMD) of some Bulge regions located
approximately along the Galactic minor axis at $b < 5^\circ$. \cite{McW10},
\cite{Nat10} and \cite{Sai11} found evidence that RC stars are split into two
components separated, on average, by 0.65\,mag along the line of sight.
\cite{McW10} argue that such a separation can be associated with an underlying
X-shaped structure of the Bulge.
\par Although the Bulge cannot be considered a single stellar system, the bulk
of its stars are mostly old with ages of the order of 10 - 12\,Gyr
\citep{Cla08,Sah06,Zoc01} thus suggesting that it is the earliest
massive component of Galaxy formed over a short time interval
\citep{McW10,Mel08,Zoc03,Ort95}. Recent empirical and theoretical investigations
support the working hypothesis that the Bulge, and in particular the inner
Bulge, was the first component of the spheroid to be formed and then it served
as a nucleus around which the rest of the Galaxy was built. This is the
so-called inside-out scenario \citep{Zoc08,Lee92}.  
According to this framework, the innermost regions of the stellar halo and of
the Bulge were formed by a rapid, dissipative collapse of low-angular momentum
material \citep{Egg62}. A high-rate of star formation, in the first
$\sim$1\,Gyr, caused a fast chemical enrichment of the interstellar medium
mainly by type\,II SNe ejecta, as confirmed by the overabundance of
$\alpha$-elements measured in Bulge stars \citep{Zoc06,McW94}. More recently,
the large spectroscopic survey of the Bulge \cite[ARGOS,][]{Fre13} using RC
stars as stellar tracers found that the metal-rich stellar component ([Fe/H] $>
-0.5$) is associated with the boxy Bulge, while the metal-poor one ([Fe/H] $<
-0.5$) is associated with the thick disk. The former component is made of two
distinct populations: the metal-poor component ([Fe/H] $\approx -0.25$) is
uniformly distributed across the selected fields while the metal-rich component
([Fe/H] $\approx 0.15$) kinematically colder and closer to the plane of the Galaxy.
The above components are considered the aftermaths of the instability-driven Bulge
formation from the pristine Galactic thin disk \citep{Nes13-1}. 
\par The observation of a radial metallicity gradient in the Bulge
\citep{Nes13-2,Utt12,Zoc08} further support the hypothesis that our Bulge is
more a boxy/peanut Bulge than a classical merger-generated Bulge
\citep{Nes13-2,Zoc08}. The Bulge GCs clusters play in this context a key role,
since they are the relics of the time when the Bulge forming instabilities took
place. In the following we briefly discuss the properties of \ngc6528 and  
four typical Baade's Window regions located around \ngc6528 and \ngc6522.  

\section{Optical CMD of the globular \ngc6528} 

One common approach to determine the properties of the Bulge is to study the
Bulge globular clusters (GCs), since this sub-system share many dynamical and
evolutionary features with the stellar population(s) of the Bulge itself
\citep{Val10,Val07,Ort95}. Among the most studied Bulge GCs there are those
located in the Baade’s Window, namely \ngc6522 and \ngc6528. 
These two clusters are projected on the same sky area and appear partially
overlapping according to recent estimates of their tidal radii
(\citeauthor{Har96} \citeyear{Har96}, as updated in 2010). \ngc6528 is an
interesting cluster, since it is among the most metal-rich GCs. High-resolution
spectroscopy, suggest for this cluster a solar metallicity and a modest
$\alpha$-element enhancement.
\cite{Car01} found $\feh = +0.07 \pm 0.01$ and a marginal $\alpha$-element
enhancement ($\afe \simeq +0.1 \pm 0.2$), while \cite{Zoc04}, by using three
giants belonging to Horizontal Branch (HB) and to the Red Giant Branch (RGB),
found $\feh = -0.1 \pm 0.2$ and $\afe \simeq +0.1 \pm 0.1$. In a more recent
investigation based on high-resolution NIR spectra \cite{Ori05} found $\feh =
-0.17 \pm 0.01$ and a higher $\alpha$-element enhancement $\afe \simeq +0.33 \pm
0.01$. According to these results \ngc6528 is an ideal laboratory not only to
constrain the $\alpha$-element enhancement in old metal-rich systems, but also
to shed new lights on the possible occurrence of an age-metallicity relation
among the most metal-rich GCs \citep{Dot11,Rak05}.
The estimate of both structural parameters and intrinsic properties for \ngc6528
have been partially hampered by the occurrence of differential reddening across
the field of view and by the high level of stellar crowding due to field stars.
The former problem can be overcome by using NIR bands, the latter one is more
complex. The use of optical-NIR color-color planes to separate field and cluster
stars \citep{Cal09,Bon10} is hampered by the fact that the metallicity
distributions of Bulge, thin disk and \ngc6528 stars peak around solar chemical
composition. Furthermore, the use of the color-color plane require precise and
deep photometry in at least three optical-NIR bands.
These are among the main reasons why current estimates of the absolute age of
\ngc6528 range from $13 \pm 2$\,Gyr \citep{Ort01} to $11 \pm 2$\,Gyr
\citep{Fel02}, to 12.6\,Gyr \citep{Mom03}.
The optical and NIR CMDs of \ngc6528 presented below come from space- and
ground-based data obtained, respectively, with \hst\ and with SOFI images. In
particular, the first dataset consists of two groups of images collected eight
years apart, between June 2002 and June 2010: the first is composed of optical
ACS/WFC images in the filters $F606W$
and $F814W$; the second comprises optical and NIR images taken, respectively,
with UVIS ($F390W, F555W, F814W$) and IR ($F110W, F160W$) channel of the
WFC3. The temporal baseline of the \hst\
dataset allowed us to compute the star proper motions in the \ngc6528 central
region corresponding to the ACS Field of View (FoV) and to split, by means of a
pure kinematic selection bona-fide cluster members from field stars
\citep{Lag14}. The \hst\ photometry was, eventually, corrected for differential
reddening using the procedure described in \cite{Mil12}.
The ground-based dataset is composed of NIR ($J$ and $K_S$) images, proprietary
data, collected with the SOFI camera at the New
Technology Telescope of ESO at La Silla, Chile. The pointings cover two FoVs
partially overlapping and slightly shifted in the South-West direction with
respect to the center of the cluster. Each pointing is roughly $5\arcmin
\times\,5\arcmin$ wide.
\par
\begin{figure*}
\resizebox{.8\textwidth}{!}{\includegraphics[clip=true]{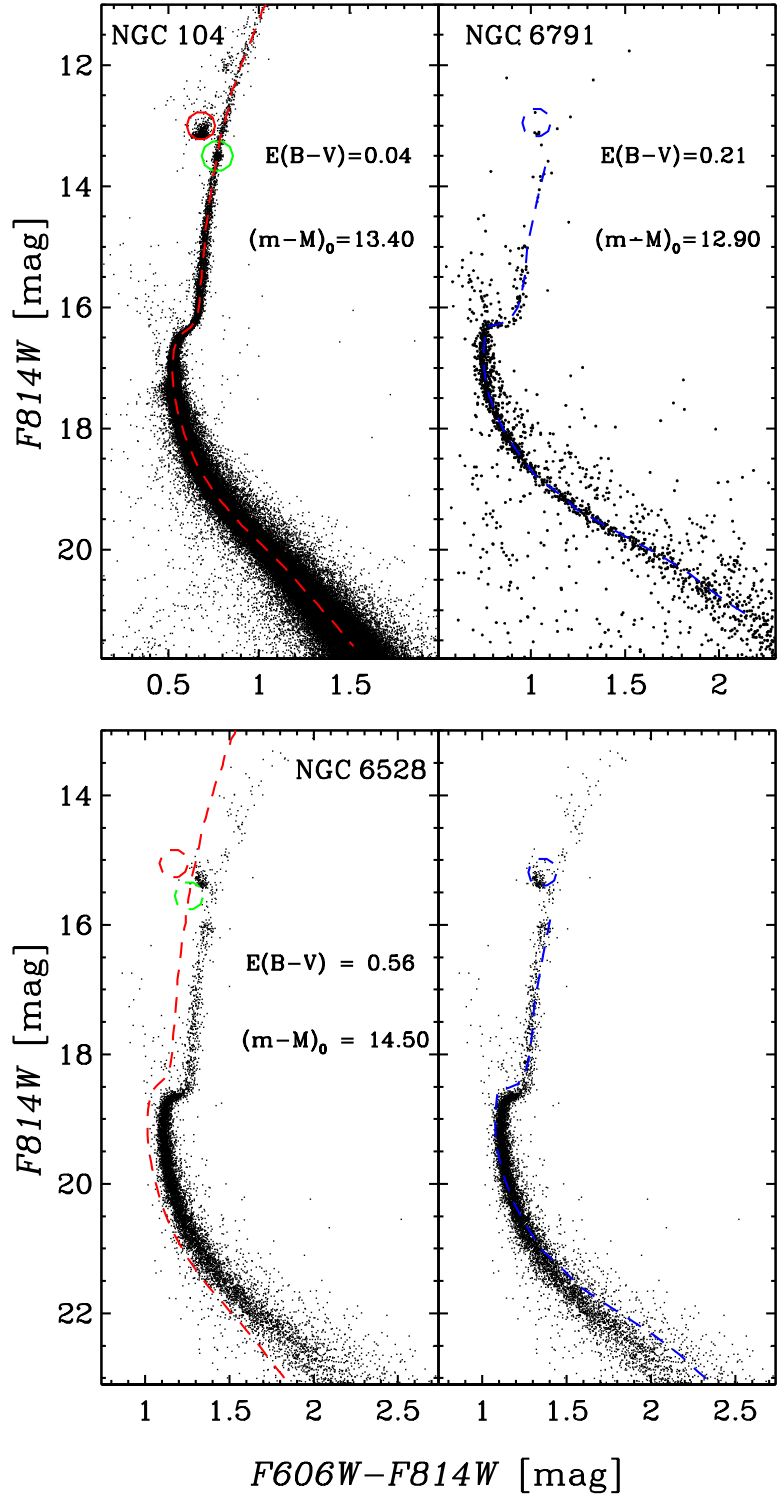}}
\centering
\caption{\footnotesize
Top-left: $F814W, F606W-F814W$ CMD of 47\,Tuc based on ACS/WFC@HST data. The
red and green circles mark, respectively, the RHB and the RGB
bump, while the dashed line represent the cluster ridgeline. Top-right: Same as
the left, but for the old, metal-rich cluster \ngc6791. The blue circle marks
the position of red clump stars.
Bottom: Comparison between the $F814W, F606W-F814W$ CMD of \ngc6528, and the
ridgelines of 47\,Tuc (left) and \ngc6791 (right). The CMD of \ngc6528 contains
only stars that, on the basis of our proper motion selection, are cluster
members and is corrected for differential reddening. The adopted true distance
modulus and reddening (\citeauthor{Har96} \citeyear{Har96}) are also labeled in
each panel. 
} 
\label{fig:fig1} 
\end{figure*}

 In order to fix the main properties of \ngc6528 we compared its CMD with
that of two metal-rich template clusters used as empirical calibrators, namely
47\,Tuc (\ngc104, $t\approx11\,{\rm Gyr}, \feh = -0.8$; \citeauthor{Van10}
\citeyear{Van10}; \citeauthor{Car09} \citeyear{Car09}) and \ngc6791
($t\approx8\,{\rm Gyr},\feh = +0.3$; \citeauthor{Bra10} \citeyear{Bra10};
\citeauthor{Boe09}\ \citeyear{Boe09}). For 47\,Tuc accurate ACS/WFC photometry
is already available ($F606W, F814W$; 47\,Tuc, \citeauthor{Cal12}
\citeyear{Cal12}, \citeauthor{And08} \citeyear{And08}); for
\ngc6791 we derived photometry in
F606W and F814W by means of the same procedure used for \ngc6528
\citep[see][]{And08}. In the upper panels of Fig.~\ref{fig:fig1} are shown the
CMDs of these two empirical calibrators (47\,Tuc at left and \ngc6791 at right),
in ACS/WFC optical bands $F606W, F814W$. For both of them we estimated the ridge
line (dashed line) and determined the magnitude and color of Red HB in 47\,Tuc
(RHB, red circle) and of RC stars in \ngc6791 (blue circle) and of the RGB Bump
in 47\,Tuc (green circle).  The Bump was not identified in \ngc6791, since it is
younger than typical globulars, and therefore with a limited number of stars
along the RGB. 
The bottom panels of Fig.~\ref{fig:fig1} show the CMD of the candidate cluster
stars for \ngc6528 in the same bands of the calibrating clusters.  The bottom
left panel shows the comparison with the ridge line, the RHB and the RGB Bump of
47\,Tuc scaled to \ngc6528 using the true distance modulus and the reddening
labeled in the figure. The comparison indicates that \ngc6528 seems more
metal-rich than 47\,Tuc. In fact the entire MS of the former is systematically
bluer than of the latter. Moreover, the RGB in \ngc6528 has a shallower slope
than the RGB of 47\,Tuc and both the RHB and the RGB bump of 47\,Tuc are
brighter and bluer than those in \ngc6528. These finding support the evidence
that the difference between the two clusters is mainly in chemical composition
rather than in age.
The comparison between \ngc6528 and \ngc6791 (bottom right panel) indicates that
the former cluster is less metal-rich than the latter. The difference between
the two depends on both chemical composition and age because the ridge line of
\ngc6791 becomes, for magnitudes fainter than $F814W \sim 21$, systematically
redder than MS stars in \ngc6528.  Moreover, the ridge line of \ngc6791 attains
colors in the TO region that are systematically brighter than main sequence
turn-off (MSTO) stars in \ngc6528 and the shape and the extent in color of the
sub-giant branch region in \ngc6791 is narrower ($F606W-F814W \sim 0.85$ {\it
vs} $F606W-F814W \sim 1.2$) compared to \ngc6528.


\section{NIR CMD of four Baade's Window fields}

\begin{figure*}
\centering
\resizebox{.75\textwidth}{!}{\includegraphics[clip=true]{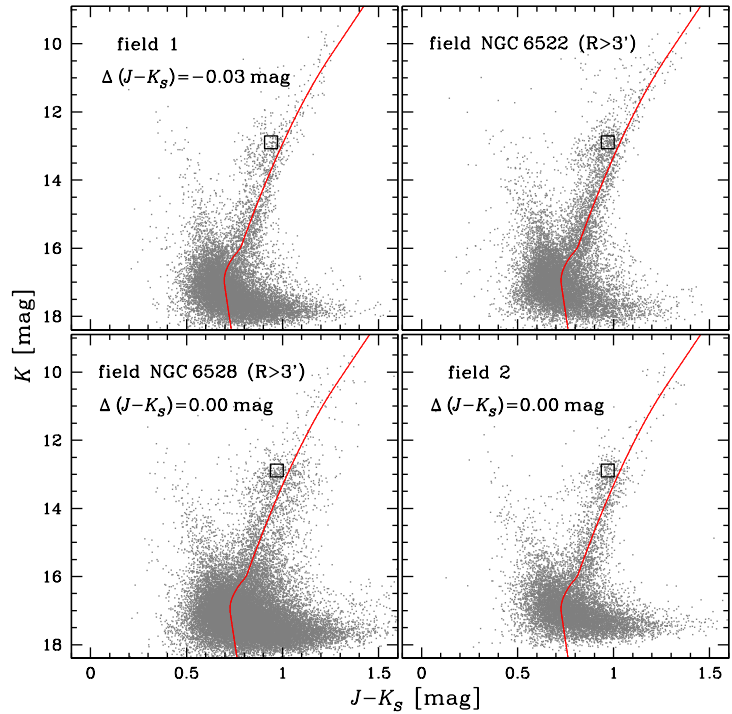}}
\caption{\footnotesize
$K_S, J -K_S$ CMDs of four Bulge fields: \ngc6528, 1, 2 and \ngc6522 (see
text for more details). The fiducial (red solid) line of the Bulge sample of the
`field \ngc6522' and the relative RHB$+$RC position (empty black square) is
overplotted on the CMD of the other fields. To account for the difference in 
mean reddening the ridge line in the top left panel was shifted in color by 
$-0.03$\,mag.\label{fig:fig2}} 
\end{figure*}

The ground based NIR images of \ngc6528 collected with SOFI at NTT are affected
by extreme crowding in the center of the cluster. We also collected homogeneous
NIR photometry of three other fields located in the Baade's Window. The PSF
photometry of the four fields was homogeneously performed \citep[see][]{Lag12}.
Fig.~\ref{fig:fig2} displays the NIR ($K_S, J-K_S$) CMDs of all the four
pointings. The CMD of the fields centered on \ngc6528 and on \ngc6522
are plotted, respectively, in the bottom-left and in the top-right panel. Note that in these
CMDs we only plotted stars located outside a circle of 3\arcmin\ from the cluster
center. This means that the above CMDs are dominated by field stars, since the
half mass radius of the two quoted clusters are 0.38 arcmin (\ngc6528) and 1
(\ngc6522) arcmin \citep{Har96}, respectively. The `field 1' (top-left) is
located slightly North-West of \ngc6528, in the outskirts of the tidal radius of
\ngc6528 and \ngc6522~\footnote{The tidal radii of \ngc6528 and \ngc6522 are:
16.57 and 16.44 arcmin \citep{Har96}.}, while the `field 2' (bottom-right) is
located southward of \ngc6522 center, in the outer part of its tidal radius. 

Data plotted in the four panels show well-defined RGBs, but current photometry
do not allow us to perform an accurate estimate of the MSTO position. Note that
the sky area covered by the fields located around the two GCs are similar, while
the sky areas of `field 1' and `field 2' are $\sim 2.5$ times smaller. 
We adopted as a reference sequence the CMD of the field located around \ngc6522.
A glance at the CMD of this field shows a well-defined concentration
of stars ranging from $K_S\simeq 13$ to $K_S\simeq 14$\,mag and $J-K_S\simeq
0.95$\,mag. This stellar overdensity is a mix between Red HB (RHB), typical of
an old stellar population (fainter, $t \gtrsim 10$\,Gyr) and RC,
typical of an intermediate-age stellar population (brighter, $1 \lesssim t <
10$\,Gyr). This means that we are dealing with the typical stellar populations
of the Galactic Bulge \citep{Sai11,McW10,Ort01}. 

We estimated the fiducial line of the `field \ngc6522' RGB and an upper limit to
the magnitude of the RHB$+$RC stars. They are plotted in the four panels of the
same figure as a red line and a black square. The mean reddening in the four
fields is quite similar, but the `field 1' for which we applied a shift in
color of $-0.03$\,mag in color. The comparison with the `field \ngc6528' shows a
large spread in color along the RGB and in the upper main sequence. Thus
suggesting that stars located in this region might be affected either by
differential reddening or by a spread in metal abundance or both.
On the other hand, the comparison with the `field 1' and `field 2' CMDs
indicates that field stellar populations are quite similar. This applies not
only to RGB and RHB$+$RC stars, but also to the blue main sequence stars located
between $K_s \sim 13$ and $K_s\sim$16\,mag and $J-K_s \sim$ 0.3 - 0.6\,mag, i.e.
the tracer of the Galactic thin disc \citep[$t\gtrsim 8$\,Gyr;][]{Cal11,McW10}.
%


Our analysis further supports the evidence that high-accuracy deep photometry is
not enough for a comprehensive study of stellar populations in the low-reddening
regions of the Bulge. The complementary use of other diagnostics (radial
velocities, proper motions, spectroscopic determination of iron content) is
mandatory to overcome the severe degeneracy (age-metallicity-reddening) of the
CMDs. 


\begin{acknowledgements} 
This work was partially supported by PRIN--INAF 2011 ``Tracing the 
formation and evolution of the Galactic halo with VST'' (PI: M. Marconi) 
and by PRIN--MIUR (2010LY5N2T) ``Chemical and dynamical evolution of the 
Milky Way and Local Group galaxies'' (PI: F. Matteucci). One of us (G.B.) 
thanks The Carnegie Observatories for support as a science visitor.
\end{acknowledgements}

\bibliographystyle{aa}

\end{document}